# High-Temperature Superconductors Explained by Pairing in Spin-Density Waves

Je Huan Koo[a] and **Kwang Chul Son**[b]*

[a]Department of Electrical and Biological Physics, Kwangwoon University,

Seoul 139-701, South Korea

[b]Graduate School of Information Contents, Kwangwoon University, Seoul 139-701, Korea

**Abstract**

The spin-density wave (SDW) can be considered as a pair of charge density waves (CDWs), one composed only of electrons with up-spins and the other only of electrons with down-spins. The high-temperature superconductivity found in cuprates and pnictides may then be ascribed to BCS-type pairing between these SDWs, which is then no longer simple Cooper pairing between independent singlet electrons but rather involves collective interaction between Cooper pairs. The pseudo-gap may also be understood to originate from this BCS-type gap of the CDW system, in which the parameters are identical to those in the original BCS scheme, except that the electron-electron interaction is multiplied by a factor $N_{CDW}$, which represents the number of electrons of the same spin direction that belong to a single CDW branch that comprises half the SDW. The superconducting gap then becomes an $s \pm d_{x^2-y^2}$-type gap. This gap may be calculated by assuming a modified BCS-like scheme that takes into account the contributions of the CDWs. These CDWs may be driven spatially into a 'checkerboard'-shaped (or striped) form of superconductivity, depending on whether the CDW is two- or one-dimensional. The origin of the nodal gap may also be ascribed to the CDWs.

*Corresponding author, e-mail:kcson@kw.ac.kr

Tel: +82-(0)2-940-8100, Fax: +82-(0)2-940-5664

## 1. Introduction

The behaviour of the low-Tc superconductors discovered by Onnes [1] may be explained satisfactorily by BCS theory [2]. In 1986 Bednorz and Müller [3] discovered high-Tc superconductivity (HTSC) by investigating the cuprates of two-dimensional electronic structures. However, the linear dependence of the resistivity on the temperature, the very high superconducting transition temperatures, and the origin of the pseudo-gap, among other things, all still require a clearer explanation if a full consensus is to be achieved regarding a theoretical model of HTSC. A number of such models have been proposed to explain the phenomenon of HTSC, including the s=3/2 hole-composite model [4,5], ferromagnetic cluster theory [6,7], spin fluctuation schemes [8-10], and the concept of resonating valence bonds [11-13]. Some theoretical approaches to HTSC are under development, both in a Heisenberg antiferromagnetic model [14] that uses the formalism of Green's function, and in the attractive Hubbard model [15], which makes use of dynamical mean field theory. Meanwhile, observations of spin-density waves (SDWs) have been made in the low-dimensional material of $(TMTSF)_2PF_6$ [16] at ambient pressures, with superconductivity being apparent at pressures above 6 kbar. However, SDWs have long been known to be incompatible with superconductivity. Recently, compounds of Fe and As with a two-dimensional structure have been studied in some depth, with investigations of relatively high superconducting transition temperatures [17] in which SDW phenomena were clearly observed [18]. Cuprate superconductors show 2-D superconductivity in the $CuO_2$ plane, and Fe-As compounds show 2-D superconductivity in the Fe-As plane. The low dimensionality of these systems is therefore linked in some way to their superconductivity. For HTSC in cuprates, both CDWs and SDWs can be observed in tandem [19-23]. At lower temperatures, it is more common to observe CDWs before SDWs [21]. However, in some cases, their simultaneous occurrence has been noted [22]. It is assumed that in such cases, the CDW phase and the BCS-type superconducting state originate in completely different channels and rarely coexist. Even in the unusual case in which the two phenomena coexist, they are in competition to expel one other [14]. The paramagnetic CDW state and the BCS-type Cooper pair state are predicted to be scarcely able to coexist. Nevertheless, the possibility cannot be excluded that antiferromagnetic CDW pairing, i.e., the existence of a CDW pair in a SDW state, may well coexist with the BCS-type superconducting state.

In the study described herein, we investigated the possibility that the SDW may be formed of pairs between up-spin and down-spin CDWs and that this antiferromagnetic CDW-pairing may coexist with superconductivity.

## 2. Model and formalism

According to BCS theory, considering the weak coupling limit, the superconducting transition temperature can be expressed as [2],

$$k_B T_c^{BCS} = 1.14 \hbar\omega \exp(\frac{1}{N(0)V}), \quad (1)$$

where $\hbar\omega$ denotes the phonon energy, $N(0)$ is the density of states at the Fermi level, and

$$V = U_{BCS} + U_c. \quad (2)$$

The phonon-mediated electron-electron interaction in the foregoing expression may be given by

$$U_{BCS} = \frac{2g^2 \hbar\omega_q}{(\varepsilon_k - \varepsilon_{k+q})^2 - (\hbar\omega_q)^2}, \quad (3)$$

where $g$ is the electron-phonon coupling constant, $\varepsilon_k$ is the kinetic energy of an electron with wavevector $k$, and $U_c$ represents the Coulomb electron-electron repulsion.

In cuprate superconductors, SDWs and CDWs are often seen to coexist [19-23]. It has previously been suggested that SDW may be composed of pairs in which the up-spin electrons form one CDW (u-CDW) and the down-spin electrons form another (d-CDW) [16]. Consider the CDWs of the SDW in which the u-CDW and the d-CDW form collective superconducting pairs.

For our CDW-pairing in the SDW [24], the wavefunction and the superconducting transition temperature are given by

$$\Phi = \prod_k \{u_k + v_k a^+_{k\uparrow} a^+_{-k\downarrow}\}^{N_{CDW}} | vac >$$

$$k_B T_c = 1.14 \hbar \omega \exp(\frac{1}{N_{CDW} N(0) V})$$

$$u_k = 1, v_k = 0 \ for \ |k| > k_F$$

$$u_k = 0, v_k = 1 \ for \ |k| < k_F$$

$|vac>$: vacuum state

$k_F$: Fermi wavevector

(4)

where $a^+_{k\uparrow}$ is the creation operator for an electron with wavevector $k$ and spin state $\uparrow$, and $N_{CDW}$ represents the number of electrons with a given spin direction in one CDW composite of the SDW.

In cuprate superconductors, we commonly observe a high-temperature structural phase transition, which may be thought of as a Peierls-Fröhlich transition accompanied by the formation of a CDW [16].

Transport in the normal state can be explained by assuming a quantum well structure associated with the pinning of the CDW, as shown in Fig. 1.

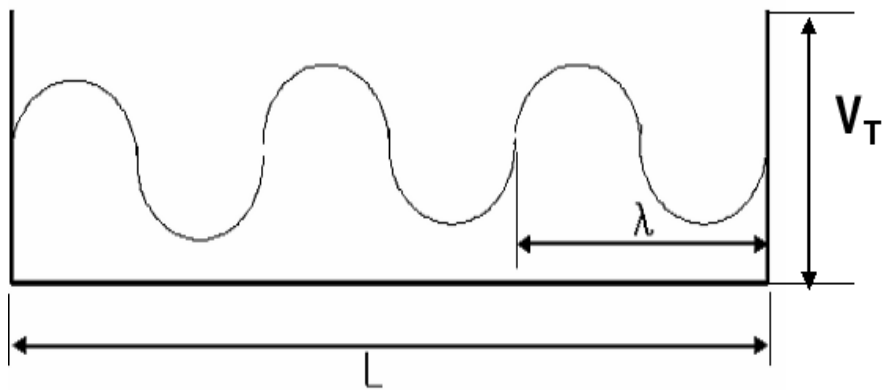

Fig. 1. The charge density wave (CDW) fitted into a typical potential well, where the confinement potential well can vary in terms of the different parameters that describe it

The current density is then given by [25]:

$$J = J_0\{\tanh[\frac{eV-eV_T}{2k_BT}]+\tanh[\frac{eV_T}{2k_BT}]\} = n_0 ev_F\{\tanh[\frac{eV-eV_T}{2k_BT}]+\tanh[\frac{eV_T}{2k_BT}]\}$$
$$= \sigma E$$
$$= \sigma \frac{V}{L} \qquad (5)$$

Here, $n_0$ is the number density of the electrons, $v_F$ is the Fermi velocity, $\sigma$ is the conductivity, $E$ is the electric field, $T$ is the temperature, $V$ is the voltage, and $V_T$ denotes the potential barrier of the confinement (as shown in Fig. 1), and $L$ is the effective size of the system. The resistivity is therefore given by

$$\rho = \frac{1}{\sigma} = \frac{1}{\frac{\partial}{\partial V}(LJ)} = \frac{2k_BT}{n_0 e^2 v_F L}\frac{1}{1-\tanh^2[\frac{-eV_T}{2k_BT}]} \qquad (6)$$

where $n_0 = N(0)\int_0^\infty f(\varepsilon)d\varepsilon = N(0)\ln(1+e^{\frac{\varepsilon_F}{k_BT}}) \simeq N(0)\varepsilon_F$.

For spin glass transitions at superconducting transition temperatures [26], the resistivity near the transition temperature can be obtained as

$$\rho = \frac{1}{\varepsilon_0}\frac{1}{\omega}\frac{1}{\tilde{\chi}''} \to \frac{1}{\varepsilon_0}\frac{1}{\omega}\frac{3S}{(S+1)(\boxed{g}eLS)^2}\frac{\beta_H}{\beta_E}k_BT - \frac{1}{\varepsilon_0}\frac{1}{\omega}\frac{3S}{(S+1)(\boxed{g}eLS)^2}\frac{\beta_H}{\beta_E}k_BT_c \; for\; T_c < T < T_{Cusp}$$

or

$$\to \frac{1}{\varepsilon_0}\frac{1}{\omega}\frac{1}{(\boxed{g}eLS)^2}\frac{\beta_H}{\beta_E}(k_BT)e^{\frac{\boxed{g}eLE}{k_BT}} - \frac{1}{\varepsilon_0}\frac{1}{\omega}\frac{1}{(\boxed{g}eLS)^2}\frac{\beta_H}{\beta_E}(k_BT_c)e^{\frac{\boxed{g}eLE}{k_BT_c}} \; for\; T_c < T < T_{Cusp}$$

(7)

where $T_c$ is the superconducting transition temperature, $T_{Cusp}$ is the cusp temperature just above $T_c$, and the other symbols correspond to those given in an earlier study [26]. In Eq. (7), the negative terms are added as a result of the temperature-independent diamagnetic contributions. In the presence of the electron-electron interaction mediated by phonon-enhanced spin-flips, $U_{sd}$, the magnetic susceptibility [27-29] of paramagnetic cuprate, is given by

$$\chi = \frac{\mu_B^2}{1 - N(0)(\tilde{V}_{sd} + \tilde{V}_c)} \qquad (8)$$

where $\mu_B$ is the Bohr magneton.

The ferromagnetic transition temperature is given by [26-28],

$$1 = N(0)(\tilde{V}_{sd} + \tilde{V}_c), \qquad (9)$$

where $\hbar\omega$ is the phonon energy, $N(0)$ is the density of states at the Fermi level,

$$\sum U_i(\cdots) \equiv \tilde{V}_i \sum (\cdots) \quad (i = c, sd) \qquad (10)$$

is the electron-electron interaction mediated by phonon-enhanced spin flips

$$U_{sd} = \frac{2g_{sd}^2 \hbar\omega_q}{(\varepsilon_k - \varepsilon_{k+q})^2 - (\hbar\omega_q)^2} \left(\frac{2s(s+1)}{3k_B T}\right)^2 n(q)n(-q), \qquad (11)$$

$g_{sd}$ is the spin-phonon coupling constant, $\varepsilon_k$ is the kinetic energy of an electron with wavevector $k$, $U_c$ is the Coulomb electron-electron repulsion, $s$ is the magnitude of spin, $k_B$ denotes the Boltzmann constant, $T$ is the temperature, and $n(q)$ represents the momentum distribution [28,29].

## 3. Results and discussions

Given that CDWs are observed in cuprates, we now consider a paramagnetic pairing between collective pairs of CDWs.

For our paramagnetic CDW pairs, the antiferromagnetic SDW pairing occurs for

$$1 = N_{CDW} N(0)(\tilde{V}_{sd} + \tilde{V}_c), \qquad (12)$$

where $N_{CDW}$ is the number of electrons in a CDW that have the preferred spin direction.

In cuprate superconductors, we consider two separate ferromagnetic sublattices, one of which is composed of an electron band from the copper lattices and the other of a hole band from the oxygen lattices. These exist as a result of the antiferromagnetic coupling between them, i.e., the antiferromagnetic interband interaction.

In BCS superconductivity, the coherence length reaches $\sim \mu m$, but in cuprate systems it is generally limited to $\sim 10 nm$.

In terms of the CDW wavelength, we obtain $N_{CDW} \sim 10$ and can therefore estimate

$$k_B T_c \sim 1.14 \times 300K \times \exp(-\frac{1}{10 \times 0.1}) \sim 100K \tag{13}$$

In the Jellium model [24], SDWs may be formed under the condition given by

$$1 - \tilde{V}(q) F(q) \leq 0 \tag{14}$$

where $\sum_k U_c (\cdots) = \tilde{V}(q) \sum_k (\cdots)$ and $F(q)$ denotes the Lindhard function.

The condition for the formation of CDWs can be expressed as

$$1 + [2U_c(q) - \tilde{V}(q)] F(q) \leq 0 \tag{15}$$

in order to find the difference of

$$2U_c(q) F(q) \approx J_{Dirac} \tag{16}$$

which corresponds to a Dirac-type exchange interaction [27]. Where this difference becomes negative, antiferromagnetic stability can be seen to bring about the formation of SDWs from the CDW pairs.

In the CDW phase, the problem of pinning becomes fairly critical [30].

To solve this, in our case we assume a simultaneous pinning of both u-CDW and d-CDW, as shown in Fig. 1, where the pinned CDW results in $L \sim 10\lambda$ and the superconducting state becomes possible.

The isotopic exponents for BCS superconductors and cuprates are respectively given by

$$\begin{aligned} \alpha_{BCS} &\approx \frac{1}{2}(1 - N(0)V) \\ \alpha_{HTSC} &\approx \frac{1}{2}(1 - N_{CDW} N(0)V) \end{aligned} \tag{17}$$

and the observed non-isotopic effect in cuprates can be explained using

$$\alpha_{HTSC} \approx \frac{1}{2}(1 - N_{CDW}N(0)V) = 0.5(1 - 10 \times 0.1) = 0 \tag{18}$$

We now consider the gap equations [31].

The gap equation for cuprates is given by

$$\Delta(T) = N_{CDW}N(0)V\Delta(T)\int_{-\hbar\omega}^{\hbar\omega} d\xi \frac{\tanh\frac{\varepsilon}{2k_BT}}{2\varepsilon}$$

$$\varepsilon = \sqrt{\xi^2 + \Delta^2(T)}$$

$$N_{CDW} = \frac{N_{CDW}(T=0)}{1 - \exp(-\frac{\hbar\omega_{CDW}}{k_BT})} \tag{19}$$

$$\omega_{CDW} = s \times 2k_F$$

where CDW denotes the bosonic mode, $\omega_{CDW}$ is the wavenumber of the CDW, and $k_F$ is the Fermi wavevector.

Using the same method used in an earlier study [31], the resulting gap equation can be rewritten as

$$\Delta(T)[1 - \frac{N_{CDW}(T=0)}{1 - \exp(-\frac{\hbar s 2k_F}{k_BT})} N(0)V \ln\frac{1.14\hbar\omega}{k_BT}]$$

$$+ \frac{N_{CDW}(T=0)}{1 - \exp(-\frac{\hbar s 2k_F}{k_BT})} \frac{N(0)V}{\pi k_B^2 T_c^2} \frac{7}{8}\zeta(3)\Delta^3(T) = 0 \tag{20}$$

where $s$ is the velocity of the electronic sound, and where $\zeta(3) = 1.2$.

If we set

$$\Delta(T) = \Delta_s(T) \pm \Delta_d(2k_F, T)[\cos^2 k_x - \cos^2 k_y], \tag{21}$$

the main equations then become

$$\Delta_s(T)[1 - N_{CDW}(T=0)N(0)V \ln\frac{1.14\hbar\omega}{k_BT}]$$

$$+ N_{CDW}(T=0)\frac{N(0)V}{\pi k_B^2 T_c^2} \frac{7}{8}\zeta(3)\Delta_s^3(T) = 0 \tag{22}$$

$$\Delta_d(2k_F,T)[-N_{CDW}(T=0)\exp(-\frac{\hbar s 2k_F}{k_B T})N(0)V\ln\frac{1.14\hbar\omega}{k_B T}]$$

$$+N_{CDW}(T=0)(1+\exp(-\frac{\hbar s 2k_F}{k_B T}))\frac{N(0)V}{\pi k_B^2 T_c^2}\frac{7}{8}\zeta(3)\{(\Delta_s(T)+\Delta_d(2k_F,T))^3(T)-\Delta_s^3(T)\} \quad (23)$$

$$+N_{CDW}(T=0)\exp(-\frac{\hbar s 2k_F}{k_B T})\frac{N(0)V}{\pi k_B^2 T_c^2}\frac{7}{8}\zeta(3)\{\Delta_s^3(T)\}=0$$

where the $k_i$ variables denote the wavevectors.

The solution is then

$$\Delta_s(T)\simeq\frac{3.5}{2}k_B T_c\sqrt{1-(\frac{k_B T}{k_B T_c})^2}$$

$$\Delta_d(2k_F,T)=0$$

or

$$\Delta_d(2k_F,T)\approx\left[0.5k_B T_c\left\langle\pm\sqrt{\frac{(\zeta(3))^2 N_{CDW}(T=0)N(0)V 2.18}{\begin{bmatrix}\{(\zeta(3))^2 N_{CDW}(T=0)N(0)V 2.18\}^2-\\ \frac{1}{k_B T_c}\{4(\zeta(3))^2 N_{CDW}(T=0)N(0)V\sqrt{1-(\frac{k_B T}{k_B T_c})^2}\times(\\ -0.416)(\ln\frac{1.14\hbar\omega}{k_B T}-\\ 3.82(\zeta(3))^2 k_B T_c N_{CDW}(T=0)N(0)V\sqrt{1-(\frac{k_B T}{k_B T_c})^2})\}\end{bmatrix}}}\right\rangle\right/\Big[ \quad (24)$$

$$(\zeta(3))^2 N_{CDW}(T=0)N(0)V\sqrt{1-(\frac{k_B T}{k_B T_c})^2}\,(1-(\frac{k_B T}{k_B T_c})^2)(-0.416)(1+\exp-\frac{\hbar s 2k_F}{k_B T})\Big]$$

where this gap function is shown in Fig. 2.

The superconducting gap becomes

$$\Delta\Rightarrow\Delta_i \quad (i=ab,c) \quad (25)$$

where these values on the ab-planes or on the c-axis are given; the total superconducting gap is also given as

$$\Delta_i^{total}\equiv\Delta_i+\Delta_i^{sd} \quad (i=ab,c),$$
$$\Delta_i^{sd}\ll\Delta_i \quad (26)$$

where $\Delta_i^{sd}$ stems from another mechanism [28], as depicted below.

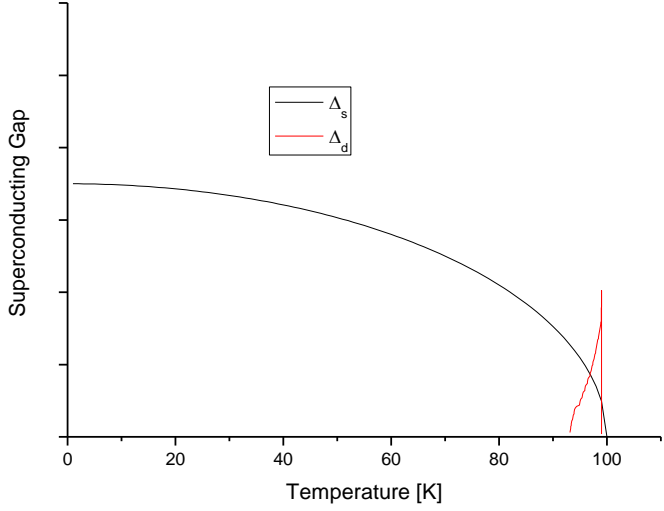

Fig. 2. The gap is shown.

The electron-electron interaction mediated by phonon-enhanced spin flips on the ab-plane and along the c-axis are rewritten as follows:

$$U_{sd} = \frac{2g_{sd}^2 \hbar \omega_q}{(\varepsilon_k - \varepsilon_{k+q})^2 - (\hbar \omega_q)^2} \left(\frac{2s(s+1)}{3k_B T}\right)^2 n(q)n(-q)$$
$$U_{sd}^c = \frac{2(g_{sd}^c)^2 \hbar \omega_q}{(\varepsilon_k - \varepsilon_{k+q})^2 - (\hbar \omega_q)^2} \left(\frac{2s(s+1)}{3k_B T}\right)^2 n(q)n(-q) \qquad (27)$$

Here, $g_{sd}$ ($g_{sd}^c$) is the spin-phonon coupling constant on the ab-planes (along the c-axis), $\varepsilon_k$ is the kinetic energy of an electron with wavevector $k$, $U_c$ is the Coulomb electron-electron repulsion, $s$ is the magnitude of the spin, $k_B$ is the Boltzmann constant, $T$ is the temperature, and $n(q)$ is the momentum distribution [28].

We choose an additional superconducting gap on the ab-plane, $\Delta_{ab}^{sd}$ as s+id determined by the form of n(q) and the superconducting gap along the c-axis, $\Delta_c^{sd}$ as s-symmetry. If we select

$$n(q) \equiv n_0 + \frac{1}{2}n_0(\cos k_x - \cos k_y)$$
$$n(-q) \equiv n_0 + \frac{1}{2}n_0(\cos k_x' - \cos k_y'), \qquad (28)$$

the average value becomes

$$<U_{sd}+U_c>=<U_{sd}^c+U_c>+<U_{sd}-U_{sd}^c>$$
$$=(U_{sd}^c+U_c)+U_{PG}(\cos k_x-\cos k_y)(\cos k_x{'}-\cos k_y{'}) \quad (29)$$
$$\Delta_{ab}^{sd}=\Delta_c^{sd}+i\Delta_{PG}(\cos k_x-\cos k_y)$$

where $\Delta_{PG}$ is the pseudo-gap from $U_{sd}-U_{sd}^c<0$ below $T_{PG}$, $U_{sd}^c\approx U_{sd}(n_0)$, and $U_{sd}(n_0)$ is only dependent on $n_0$ only if $g_{sd}=g_{sd}^c$. The additional gaps are given as

$$\Delta_{ab}^{sd}=\sum_{k'}|U_{sd}+U_c|(k,k')\frac{\Delta_{ab}^{sd}(k')}{2W_{k'}}\tanh(\frac{W_{k'}}{2k_BT})$$
$$\Delta_c^{sd}=\sum_{k'}|U_{sd}^c+U_c|(k,k')\frac{\Delta_c^{sd}(k')}{2W_{k'}}\tanh(\frac{W_{k'}}{2k_BT}) \quad (30)$$
$$\Delta_{PG}=|U_{PG}|\int\frac{d^2\vec{k}}{(2\pi)^2}\frac{\Delta_{PG}(\cos k_x-\cos k_y)^2}{2W_{k'}}\tanh(\frac{W_k}{2k_BT})$$
$$W_k^2=\xi_k^2+(\Delta_c^{sd})^2+\Delta_{PG}^2(\cos k_x-\cos k_y)^2$$

$$\xi_k=\varepsilon_k^{sd}-\varepsilon_F^{sd}, \varepsilon_i^{sd}=\varepsilon_i-\varepsilon_{sd}\ (i=k,F), U_{PG}=\frac{1}{4}U_{sd}^c,$$

where $\varepsilon_{sd}=\frac{2J_{sd,0}^2N_0s(s+1)}{3k_BT}\sum_{k,\sigma}\langle a_{k,\sigma}^+a_{k,\sigma}\rangle$ is given in the aforementioned study [28].

At the onset temperature of the pseudo-gap $T_{PG}$, this becomes

$$k_BT_{PG}\approx 1.14\hbar\omega\exp\frac{-1}{4N(\varepsilon_F)|U_{PG}|}. \quad (31)$$

The pseudo-gap is shown in Fig. 3.

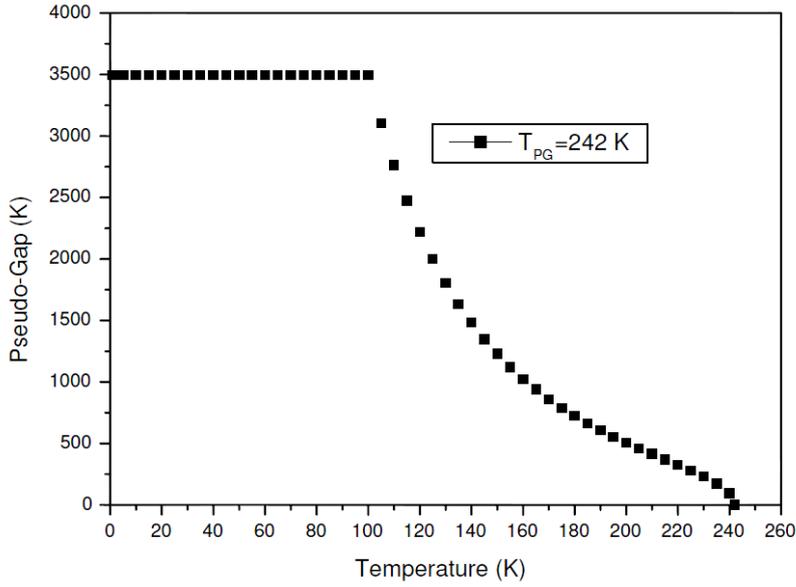

Fig. 3. The pseudo-gap is plotted for given parameters as $U_c = 3.75 \times 10000\ K, U_{sd}^c(T_c) = -4.4 \times 10000\ K, T_c = 100\ K, T_{PG} = 242\ K$.

## 4. Conclusions

2-D structures of copper oxide superconductor systems undergo a structural phase transition when the temperature is lowered from the metallic state, when the Peierls-Fröhlich transition [16] brings about the formation of CDWs. In the SDW phase, if the BCS-type electron-electron interaction as expressed in Eq. (2) becomes attractive, the superconducting states are brought about in accordance with Eq. (4). From Eqs. (21)-(24), two-dimensional (one-dimensional) CDWs can drive in spatial terms a checkerboard (striped) type of superconductivity [32]. A nodal superconducting gap may occur along the CDW, but this arises in no other direction apart from the direction of the CDW. The superconducting gap originates mainly from the CDW and from phonon-enhanced spin flips. The pseudo-gap serves as a charge gap from the difference between the electron-electron interaction mediated by phonon-enhanced spin flips on the ab planes, while along the c-axis is only a type of anisotropic gap, that is, dimensional crossover from 3D to 2D. In Eqs. (9) and (30), the denominator can be positive in the case of interband interactions between Cu and O but negative for intraband interactions.

## Acknowledgement

This work was supported by the research grant of Kwangwoon University in 2013.

**Figure caption**

Fig. 1. The charge density wave (CDW) fitted into a typical potential well, where the confinement potential well can vary in terms of the different parameters that describe it

Fig. 2. The gap is shown.

Fig. 3. The pseudo-gap is plotted for given parameters as $U_c = 3.75 \times 10000\ K, U_{sd}^c(T_c) = -4.4 \times 10000\ K, T_c = 100\ K, T_{PG} = 242\ K.$

15